\documentclass[10pt]{article}
\usepackage{graphicx}
\begin{document}

\title{SPONTANEOUS PION EMISSION DURING FISSION A NEW NUCLEAR RADIOACTIVITY}
\author{D. B. Ion$^{1}$, Reveica Ion-Mihai$^{2}$, M. L. Ion$^{2}$ and \\
Adriana I. Sandru$^{1}$ \\\\
$^{1}$ {\it National Institute for Physics and Nuclear Engineering,}\\
{\it IFIN-HH}, Bucharest, Romania\\
$^{2}$ {\it University of Bucharest, Department of Atomic and} \\
{\it Nuclear Physics}, Bucharest, Romania}

\maketitle

\begin{abstract}
In this paper a short review of the theoretical problems of the pionic
radioactivity is presented.  The essential experimental results obtained
in the 18 years of existence of the nuclear pionic radioactivity are reviewed.
Moreover, using the recent results on the spontaneous fission half lives $T_{SF}$
of the heavy nuclei with $Z\geq100$  new predictions on the pionic yields
in the region of superheavy elements are established.
\end{abstract}

\section{INTRODUCTION}

In 1935 Japanese physicist Hideki Yukawa predicted [1] the existence of a
particle - a meson - with a mass about 200 times that of the electron, which
mediated the nuclear forces. At first, the muon, discovered in 1937, looked
like a good candidate for the meson, but soon it became clear that the
muon's properties did not match those predicted by Yukawa theory. Then, in 1947,
Lates, Occhialini and Powell discovered [2] the $\pi $ -meson, or pion,
which did have the predicted meson properties. Today, we know that the muon
is an elementary particle [3], cousin to the electron, -leptons and
neutrinos, and that mesons ($\pi $, $K$, $\eta $ , etc.), including the pion,
are combinations of quark and an antiquark. Everybody know that the
discovery of -meson was an very important finding since this was in fact a
fundamental step in the understanding the subatomic world. It heralded the
beginning of a deep revision of the physical concepts on the structure of
matter. However, who can tell us what exactly are those mesons? Also, it is
well known that the pion as lightest of mesons has a finite size, with a
mean charge radius of 0.66 fm, and that the long range part of the
nucleon-nucleon forces necessarily arises from one-pion-exchange.

The traditional picture of the nucleus as a collection of neutrons and
protons bound together via the strong force has proven remarkable successful
in understanding a rich variety of nuclear properties. However, it is a well
grounded achievement of modern nuclear physics that not only nucleons are
relevant in the study of nuclear dynamics but that pions and the baryonic
resonances like $\Delta$'s and N* play an important role too [4]. So, when the
nucleus is explored at short distance scales the presence of short lived
subatomic particles, such as the pion and delta, are revealed as nuclear
constituents. At even shorter distance scales the basic building blocks of
matter, the quarks and gluons, are also revealed as nuclear constituents.
The role of pions, deltas, quarks and gluons in the structure of nuclei is
one frontier of modern nuclear physics. This modern picture of the nucleus
bring us to the idea to search for new kind of natural radioactivities such
as: ($\pi$, $\mu$, $\Delta$ ,$N*$ etc.)-emission during the nuclear fission
in the region of heavy and
superheavy nuclei. Moreover, new mode of nuclear fission such as: deltonic
fission an hyperfission, was also suggested and investigated. So, in 1985,
D. B. Ion, M. Ivascu and R. Ion-Mihai initiated [5] the investigation of the
nuclear spontaneous pion emission as a new possible nuclear radioactivity
called nuclear pionic radioactivity (NPIR) with possible essential
contributions to the instability of heavy and superheavy nuclei.

In this paper we present a short review not only of the main theoretical
problems of the NPIR but also of the essential experimental results obtained
in these 18 years of existence of the nuclear pionic radioactivity.
Moreover, in this paper by using the recent results on the spontaneous
fission half lives $T_{SF}$ of the heavy nuclei with $Z>100$ we present
new prediction on the pionic yields in the region of superheavy elements.

\section{FISSION-LIKE MODEL FOR PIONIC RADIOACTIVITY}

The nuclear pionic radioactivity of a parent nucleus (A,Z) can be considered
as an inclusive reaction of form:

\begin{equation}
(A,Z)\rightarrow \pi +X  \label{1}
\end{equation}
where X denotes any configuration of final particles (fragments, light
neutral and charged particles, etc.) which accompany emission process. The
inclusive NPIR is in fact a sum of all exclusive nuclear reactions allowed
by the conservation laws in which a pion can be emitted by a nucleus from
its ground state. The most important exclusive reactions which give the
essential contribution to the inclusive NPIR (1) are the spontaneous pion
emission accompanied by two body fission::
\begin{equation}
_{Z}^{A}X\rightarrow \pi +_{Z_{1}}^{A_{1}}X+_{Z_{2}}^{A_{2}}X  \label{2}
\end{equation}
where

\begin{equation}
\begin{tabular}{l}
$A=A_{1}+A_{2}$ \\
$Z=Z_{1}+Z_{2}+Z_{\pi }$%
\end{tabular}
\label{3}
\end{equation}

Hence, the NPIR is an extremely complex coherent reaction in which we are
dealing with a spontaneous pion emission accompanied by a rearrangement of
the parent nucleus in two or many final nuclei. Charged pions as well as
neutral pions can be emitted during two body or many body fission of parent
nucleus.

The energy released in an exclusive nuclear reaction (1) is given by

\begin{equation}
Q_{\pi F}=Q_{SF}-m_{\pi }  \label{4}
\end{equation}
where $Q_{SF}$ is the energy released during spontaneous fission

\begin{equation}
Q_{SF}=m(Z,A)-m(Z_{1},A_{1})-m(Z_{2},A_{2})  \label{5}
\end{equation}

Then the energy condition for the pionic radioactivity channel is

\begin{equation}
Q_{\pi F}>0  \label{6}
\end{equation}
or
\begin{equation}
Q_{SF}>m_{\pi }  \label{7}
\end{equation}

Other interesting new natural radioactivities predicted at IFIN-Bucharest
are as follows

\begin{itemize}
\item \textbf{Muonic radioactivity} [25] is a nuclear process in which the lepton pair $%
\left( \mu ^{\pm },\upsilon _{\mu }\right) $ is emitted during two or many body
fission of the parent nucleus. For the the nuclear spontaneous emission of
$\left( \mu ^{\pm },\upsilon _{\mu }\right) $ in binary fission we can write

\begin{equation}
_{Z}^{A}X\rightarrow \mu ^{\pm }+\nu _{\mu
}+_{Z_{1}}^{A_{1}}X+_{Z_{2}}^{A_{2}}X  \label{8}
\end{equation}

\begin{equation}
\begin{tabular}{l}
$A=A_{1}+A_{2}$ \\
$Z=Z_{1}+Z_{2}+Z_{\mu }$%
\end{tabular}
\label{9}
\end{equation}

\item \textbf{Lambdonic radioactivity} [26] is just a nuclear
reaction of form

\begin{equation}
_{Z}^{A}X\rightarrow \Lambda ^{0}+_{Z_{1}}^{A_{1}}X+_{Z_{2}}^{A_{2}}X
\label{10}
\end{equation}

\begin{equation}
\begin{tabular}{l}
$A=A_{1}+A_{2}$ \\
$Z=Z_{1}+Z_{2}$%
\end{tabular}
\label{11}
\end{equation}

\item \textbf{Hyperonic radioactivities} $\left( \Sigma
^{-},\Sigma ^{0},\Sigma ^{+}\right) $ are also the possible
nuclear decays of form

\begin{equation}
_{Z}^{A}X\rightarrow \Sigma ^{\pm ,0}+_{Z_{1}}^{A_{1}}X+_{Z_{2}}^{A_{2}}X
\label{12}
\end{equation}

\begin{equation}
\begin{tabular}{l}
$A=A_{1}+A_{2}$ \\
$Z=Z_{1}+Z_{2}+Z_{\Sigma }$%
\end{tabular}
\label{13}
\end{equation}

\item \textbf{Hyperfission or weak fission} [27] is a fission
process in which in one of fragment is hypernucleus:

\begin{equation}
_{Z}^{A}X\rightarrow _{Z_{1}}^{A_{1}}X+_{Z_{2}}^{A_{2}}X_{\gamma }
\label{14}
\end{equation}

\begin{equation}
\begin{tabular}{l}
$A=A_{1}+A_{2}$ \\
$Z=Z_{1}+Z_{2}$%
\end{tabular}
\label{15}
\end{equation}

\item \textbf{Deltonic fission} [28] is also a fission process in
which in one of fragment contain a delta resonance:

\begin{equation}
_{Z}^{A}X\rightarrow _{Z_{1}}^{A_{1}}X+_{Z_{2}}^{A_{2}}X_{\Delta }
\label{16}
\end{equation}

\begin{equation}
\begin{tabular}{l}
$A=A_{1}+A_{2}$ \\
$Z=Z_{1}+Z_{2}$%
\end{tabular}
\label{17}
\end{equation}

\end{itemize}
A fission-like model (see Ref.[5]) for the pionic radioactivity
was regarded as a first stage in the development of an approximate
theory of this new phenomenon that takes into account the
essential degree of freedom of the system: $\pi$-fissility,
$\pi$-fission barrier height, etc.

Therefore, let us consider

\begin{equation}
E_{C}^{\pi F}\left( 0\right) =E_{C}^{0}-\alpha m_{\pi }  \label{18}
\end{equation}

\begin{equation}
E_{S}^{\pi F}\left( 0\right) =E_{S}^{0}-\left( 1-\alpha \right) m_{\pi }
\label{19}
\end{equation}
where $E_{C}^{0}$ and $E_{S}^{0}$ are the usual Coulomb energy and surface
energy of the parent nucleus given by

\begin{equation}
E_{C}^{0}=\gamma Z^{2}/A^{1/3}  \label{20}
\end{equation}

\begin{equation}
E_{S}^{0}=\beta A^{2/3}  \label{21}
\end{equation}
with $\beta =$MeV and $\gamma =$0.71 MeV.$\alpha $ is a parameter
defined so that $\alpha m_{\pi }$ and $\left( 1-\alpha \right)
m_{\pi }$ are the Coulombian and nuclear contributions to the pion
mass. For $\alpha =1$, the entire pion mass is obtained only from
Coulomb energy of the parent nucleus.
So, by analogy with binary fission was introduced the pionic fissility $%
X_{\pi F}$ which is given by

\begin{equation}
X_{\pi F}^{\left( \alpha \right) }=\frac{E_{C}^{\pi F}\left( 0\right) }{%
2E_{S}^{\pi F}\left( 0\right) }=\frac{E_{C}^{0}-\alpha m_{\alpha }}{%
E_{S}^{0}-\left( 1-\alpha \right) m_{\pi }}  \label{22}
\end{equation}
or

\begin{equation}
\left( \frac{Z^{2}}{A}\right) _{\pi F}=\frac{Z^{2}}{A}-\frac{m_{\pi }}{%
\gamma A^{2/3}}\frac{\alpha -\left( 1-\alpha \right) E_{C}^{0}/E_{S}^{0}}{%
1-\left( 1-\alpha \right) m_{\pi }/E_{S}^{0}}  \label{23}
\end{equation}

In the particular case $\alpha=1$ we have

\begin{equation}
X_{\pi F}=X_{SF}-\frac{m_{\pi }}{2E_{S}^{0}}  \label{24}
\end{equation}

We note that the definitions (22)-(24) are valid only with the
constraints

\begin{equation}
E_{C}^{\pi F}\left( 0\right) +E_{S}^{\pi F}\left( 0\right) +m_{\pi
}=E_{C}^{0}+E_{S}^{0}  \label{25}
\end{equation}
while the total variation $\Delta E^{\pi F}$of parent nucleus at small
deformation of type $\varepsilon P_{2}\left( \cos \theta \right) $is given by

\begin{equation}
\Delta E^{\pi F}=\Delta E_{C}^{\pi F}+\Delta E_{S}^{\pi F}=\frac{\varepsilon
^{2}}{5}\left[ 2E_{S}^{\pi F}\left( 0\right) -E_{C}^{\pi F}\left( 0\right)
\right] =\frac{2\varepsilon ^{2}E_{S}^{\pi F}\left( 0\right) }{5}\left[
1-X_{\pi F}^{\left( \alpha \right) }\right]  \label{26}
\end{equation}

In Fig. 2 we presented the regions from the plane (A,Z) in which some parent
nuclei are able to emit spontaneously pions during the nuclear fission

Therefore, according to Fig. 2, we have the following important regions:
\begin{itemize}
\item SHE (super heavy elements)-region, indicated by white
circles, where $X_{\pi F}>1$ and $Q_{\pi}>0$, all the nuclei are
able to emit spontaneously pions during the SF since no pion
fission barrier exists.

\item HE (heavy elements)-zone indicated by signs plus (+++),
corresponding to $X_{\pi F}>1$ and $Q_{\pi}>0$, where all the
nuclei can emit spontaneously pion only by quantum tunneling of
the pionic fission barrier.

\item E-region, indicated by signs minus (---) where the
spontaneous pion emission is energetically interdicted since
$Q_{\pi}<0$.
\end{itemize}
The dynamical thresholds for the pionic fission are obtained just as in
fission case by using the substitution:

\begin{equation}
X_{SF}\rightarrow X_{\pi F};E_{C}^{0}\rightarrow E_{C}^{\pi F}\left(
0\right) ,E_{S}^{0}\rightarrow E_{S}^{\pi F}\left( 0\right)   \label{27}
\end{equation}
So, by analogy with binary fission the barrier height for the
pionic fission in a liquid drop model can be written as:

\begin{equation}
E^{\pi F}\left( LD\right) =E_{S}\left( 0\right) \left[ 0.73\left(
1-X_{\pi F}\right) ^{3}-0.33\left( 1-X_{\pi F}\right)
^{4}+1.92\left( 1-X_{\pi F}\right) ^{5}-0.21\left( 1-X_{\pi
F}\right) ^{6}\right]   \label{28}
\end{equation}
while the nuclear configuration at the sadle point is given by

\begin{equation}
R\left( \theta \right) =\frac{R_{0}}{\lambda }\left[ 1+\varepsilon
_{2}P_{2}\left( \cos \theta \right) +\varepsilon _{4}P_{4}\left( \cos \theta
\right) +\varepsilon _{6}P_{6}\left( \cos \theta \right) +...\right]
\label{29}
\end{equation}
where

\begin{equation}
\begin{tabular}{l}
$\varepsilon _{2}=2.33\left( 1-X_{\pi F}\right) -1.23\left(
1-X_{\pi F}\right) ^{2}+9.50\left( 1-X_{\pi F}\right)
^{3}-8.05\left( 1-X_{\pi
F}\right) ^{4}+...$ \\
$\varepsilon _{4}=1.98\left( 1-X_{\pi F}\right) ^{2}-1.70\left(
1-X_{\pi
F}\right) ^{3}+17.74\left( 1-X_{\pi F}\right) ^{4}+...$ \\
$\varepsilon _{6}=-0.95\left( 1-X_{\pi F}\right) +...$%
\end{tabular}
\label{30}
\end{equation}
$R_{0}$ is the spherical radius and $\lambda$ is a scale factor
just as in binary spontaneous fission.

In Fig. 4 and 5 we present the values of E$^{\pi F}\left(
LD\right) $ as well as the nuclear configuration at the saddle
point compared with those from fission (F) or hyperfission (HF).

True barrier height for the pionic fission is
\begin{equation}
E^{\pi F}=E^{\pi F}\left( LD\right) -\Delta E_{shell}^{\pi
F}\left( s.p.\right) -\Delta E_{shell}\left( g.s.\right)
\label{31}
\end{equation}
where and $\Delta E_{shell}^{\pi F}\left( s.p\right) $and $\Delta
E_{shell}\left( g.s\right) $ are correction due to shell effect at
saddle point and ground state, respectively.

Next, the essential ingredients used for the estimation of the
relative pionic yields $\frac{\Gamma _{\pi F}}{\Gamma _{SF}}$ are
as follows:

\begin{itemize}
\item the variables $\theta $ and $\theta _{\pi F}$ defined by

\begin{equation}
\theta =\frac{Z^{2}}{A}-37,5  \label{32}
\end{equation}

\begin{equation}
\theta _{\pi F}=\left( \frac{Z^{2}}{A}\right) _{\pi F}-37,5=\theta -\frac{%
m_{\pi }}{\gamma A^{2/3}}\frac{\alpha -\left( 1-\alpha \right)
E_{C}^{0}/E_{S}^{0}}{1-\left( 1-\alpha \right) m_{\pi }/E_{S}^{0}}
\label{33}
\end{equation}

\item the real function $\tau \left( \theta \right) $ defined for
spontaneous fission (SF) by

\begin{equation}
\tau \left( \theta \right) =\log _{10}T_{SF}\left( \theta \right) +(5-\theta
)\delta M=a_{0}+a_{1}\theta +a_{2}\theta ^{2}+..  \label{34}
\end{equation}

\item the scaling property:

\begin{equation}
\tau (\theta )=\log _{10}T_{\pi F}\left( \theta _{\pi F}\right) +(5-\theta
_{\pi F})\delta M=a_{0}+a_{1}\theta _{\pi F}+a_{2}\theta _{\pi F}^{2}+..
\label{35}
\end{equation}
where $\delta M$ is defined in Ref. [5].
\end{itemize}

By fit were obtained the following important results:

\begin{equation}
\tau (\theta )=19.70-6.13\theta  \label{36}
\end{equation}
for even-even parent nuclei

and

\begin{equation}
\tau (\theta )=28.21-7.32\theta  \label{37}
\end{equation}
for parent nuclei with A-odd.

For example in Fig. 6 we presented the experimental values of the $\tau
(\theta )$\ compared with the fitted curve (37).

Finally, by using (36)-(37) we obtain the following important prediction

\begin{equation}
\frac{\Gamma _{\pi }}{\Gamma _{SF}}=\left[ \frac{T_{SF}}{T_{SF}^{C}}\right]
^{\delta _{\alpha }(A,Z)}  \label{38}
\end{equation}

where

\begin{equation}
\delta _{\alpha }(A,Z)=\frac{\Delta \theta _{\alpha }}{\theta -5}  \label{39}
\end{equation}

and

\begin{equation}
\Delta \theta _{\alpha }=\frac{m_{\pi }}{\gamma A^{2/3}}\frac{\alpha
-(1-\alpha )E_{C}^{0}/E_{S}^{0}}{1-\left( 1-\alpha \right) m_{\pi }/E_{S}^{0}%
}  \label{40}
\end{equation}

\begin{eqnarray}
E_{C}^{0} &=&0,7053Z^{2}A^{-1/3}(MeV)  \label{41} \\
&&and  \nonumber \\
E_{S}^{0} &=&17,80A^{2/3}(MeV)  \nonumber
\end{eqnarray}

\begin{equation}
T_{SF}^{C}=10^{-10,95}(yr)  \label{42}
\end{equation}
for even-even parent nuclei, and

\begin{equation}
T_{SF}^{C}=10^{-8,39}(yr)  \label{43}
\end{equation}
for A-odd parent nuclei.

Here, as in Ref. [31] we also adopt the notions of
\textit{SF-nuclide}, \textit{$\pi F$-nuclide}, and
\textit{T-transition nuclide}, as being those parent nuclei
characterized by : $\frac{\Gamma _{\pi ^{0,\pm }}}{\Gamma _{SF}}<1$, $\frac{%
\Gamma _{\pi ^{0,\pm }}}{\Gamma _{SF}}>1$\ and yield $\Gamma _{\pi ^{0,\pm
}}=\Gamma _{SF}$\ , respectively, with their characteristic features
presented in Table 1.

The predicted values of $\frac{\Gamma _{\pi ^{0,\pm }}}{\Gamma
_{SF}}$\ calculated with (38)-(40) for $\alpha =1$\ and $\alpha
=0.75$\ , for even-even and A-odd parent nuclei, are presented in
Table 2a,b.

\section{EXPERIMENTAL SEARCH FOR NUCLEAR PIONIC RADIOACTIVITY}

\subsection{Experimental limits on NPIR-yield}

The pionic radioactivity was experimentally investigated by many authors
[13]-[24]. A short review of the results obtained on the spontaneous NPIR
yields is presented in Table 3.

\subsection{Recent results from LIPPE-Obninsk (Russia)}

Recently Khryachkov et al. [24] presented a new high performance
spectrometer for investigation of ternary nuclear fission. The
measured characteristics of this spectrometer allow for its
successful use in studies of ternary fission with the emission of
$\alpha $-particles, tritons, and protons as well as in the search
for exotic nuclear fission accompanied by the emission of charged
mesons ($\pi ^{\pm },\mu ^{\pm })$. This new spectrometer was
tested with a reaction of spontaneous $^{252}Cf$ ternary fission.
This choice was determined by the fact that this reaction is well
studied and the available
data can be employed for the performance check of this facility. So, a $%
^{252}Cf$ layer 5mm in diameter with an activity of 15 fissions/s
was placed inside the spectrometer. The measurements were carried
out continuously 1.5 months. A 2D spectrum of the scintillator
signals obtained in coincidence with fragments is shown here in
Fig.8 (see Fig. 9 in Ref. [24]).

Analyzing spectrum in Fig. 8 has shown that, \textit{in the space
between the electron and proton regions there are events that
cannot be explained by an external background or as effect of
neutrons}. A detailed analysis of the parameter SF does not allow
us to assign these events to protons or electrons. We estimate
these events as an indication of the possibility of existence of a
ternary nuclear decay with the participation of charged mesons.
Subsequently, it is suggested to study the meson emission during
the spontaneous fission on this new facility which has a number of
substantial advantages.

\section{SUMMARY AND CONCLUSIONS}

The experimental and theoretical results obtained on the NPIR can be
summarized as follows:

1. The best experimental limit for $\pi$-yields has been reported
for $^{252}Cf$ by Bellini et al. [19]. They reached an upper limit
of $3\cdot10^{-13}$, a value close to the theoretical prediction
[8].

2. The unusual background, observed by Wild et al.[33] in ($\Delta
E-E$)-energy region below that characteristic for long range alpha
emission from $^{257}Fm$ was interpreted by Ion, Bucurescu and
Ion-Mihai [9] as being produced by negative pions emitted
spontaneously by $^{257}Fm$. Then, they inferred value of the
pionic yield is: $\Gamma_{\pi}/\Gamma_{SF} =(1.2\pm 0.2)\cdot
10^{-3}$ for $\pi^{-}$/fission. In a similar way, Janko and
Povinec [22], obtained the yield : $\Gamma_{\pi}/\Gamma_{SF}
=(7\pm )\cdot 10^{-5}$ for $\pi^{+}$/fission.

3. The supergiant radiohalos (SGH), discovered by Grady and Walker
[34] and Laemmlein [35] are interpreted [12] as being the
$\pi^{-}$-radiohalos and $\pi^{+}$-radiohalos, respectively. Hence
these radiohalos are experimental signatures of the integral
record in time of the natural pionic radioactivity from
radioactive inclusions in ancient minerals.

4. A new interpretation of the experimental bimodal fission is
obtained in terms of the unitarity diagrams. is obtained in Ref.
[29] So, the presence of the symmetric mode in the fragment
mass-distribution at transfermium nuclei can be interpreted as
experimental signature of the pionic radioactivity. Hence, it is
expected that this new degree of nuclear instability can becomes
dominant at SHE-nuclides. This idea is illustrated not only by the
symmetrization of fragment mass-yields from SF of the heavy nuclei
with $98\leq Z\leq 104$ experimentally observed in Fig. 7 but also
by high values of $\Gamma_{\pi}/\Gamma_{SF}$-yields obtained by
using the theoretical prediction (38)-(39).

5. The nuclides $No$ with $A=242-250$ are expected to be all $\pi
F$-nuclides while the $No$-isotopes with $A\geq 252$ are all
SF-nuclides (see Table 2a). High pionic yields are obtained for
$^{250}No$ (6.67) as well as for $^{258}No$ ($7.1\cdot 10^{-2}$)
and $ ^{262}No$ ($1.2\cdot 10^{-2}$).

6. The nuclides $Lr$ and $Rf$ are all predicted to be SF-nuclides
(see Table 2a).

7. The nuclides $^{258}Sg$ ($1.3\cdot 10^{7}$), $^{259}Sg$
($6.1\cdot 10^{6}$), $^{260}Sg$ ($1.2\cdot 10^{8}$), $^{261}Sg$
($1.3\cdot 10^{11}$), $^{263}Sg$ ($5.2\cdot 10^27$) as well as
$^{264}Hs$ ($1.2\cdot 10^{2}$) all are predicted to be $\pi
F$-nuclides.

Finally, we note that dedicated experiments, using nuclei with theoretically
predicted high pionic yields (e.g. $^{258}Fm$, $^{259}Fm$, $^{258}No$, $%
^{260}No$, $^{254}Rf$, $Sg$ and $^{264}Hs$ ($1.2\cdot 10^{2}$)),
are desired, since the discovery of the nuclear pionic
radioactivity is of fundamental importance not only in nuclear
science (e. g., for clarification of high instability of SHE-
nuclei) but also for geochemistry and cosmology.

\vspace{1cm} 

\textbf{Table 1}: Definitions and characteristic features of (SF, $\pi F$ , T)-nuclides.

\noindent \begin{tabular}{|l|l|l|l|l|l|}
\hline
Type of & Pionic & Mass & Half Lives & Half Lives  & Dominant \\
nuclide & Yield & Distribution & $Z^{2}/A<42.5$ &
$Z^{2}/A>42.5$ & Unitarity  \\ 
& & (MD) & & &Diagram \\ \hline
SF-nuclide &
$\frac{\Gamma _{\pi ^{0,\pm }}}{\Gamma _{SF}}<1$ & Asymmetric MD
& $T_{SF}>T_{SF}^{C}$ & $T_{SF}<T_{SF}^{C}$ & $D_{SF}$ \\
T-nuclide & $\frac{\Gamma _{\pi ^{0,\pm }}}{\Gamma _{SF}}=1$ & Bimodal MD & $%
T_{SF}=T_{SF}^{C}$ & $T_{SF}=T_{SF}^{C}$ & $D_{\pi F}+D_{SF}$ \\
$\pi F-$nuclide & $\frac{\Gamma _{\pi ^{0,\pm }}}{\Gamma _{SF}}>1$ &
Symmetric MD & $T_{SF}<T_{SF}^{C}$ & $T_{SF}>T_{SF}^{C}$ & a new phase \\ 
&&&&& $D_{\pi F}$ n$\Longleftrightarrow p$ \\ \hline
\end{tabular}

\vspace{1cm} 

\textbf{Table 2a}: Theoretical predictions of the $\frac{\Gamma _{\pi
^{0,\pm }}}{\Gamma _{SF}}$\ yields for heavy parent nuclei
with Z between Z=100 and Z=108, obtained for $\alpha =1$\ by using Eqs.
(38)-(40) and $T_{SF}$ from Ref. [32].

\noindent $
\begin{array}[b]{|l|l|l|l|l|l|l|l|} \hline
& Z & A & T_{SF} & T_{SF}/T_{SF}^{C} & \Gamma _{\pi ^{-}}/\Gamma _{SF} &
\Gamma _{\pi ^{0}}/\Gamma _{SF} & \Gamma _{\pi ^{+}}/\Gamma _{SF} \\ \hline
Fm & 100 & 242 & 2.535\cdot 10^{-11} & 2.259\cdot 10^{+00} & 3.55\cdot 10^{-02} & 3.30\cdot 10^{-02} & 2.43\cdot 10^{-02} \\
& 100 & 244 & 1.046\cdot 10^{-10} & 9.320\cdot 10^{+00} & 8.59\cdot 10^{-04} & 7.35\cdot 10^{-04} & 3.85\cdot 10^{-04} \\
& 100 & 245 & 1.255\cdot 10^{-04} & 3.080\cdot 10^{+04} & 1.77\cdot 10^{-13} & 9.32\cdot 10^{-14} & 6.33\cdot 10^{-15} \\
& 100 & 246 & 4.753\cdot 10^{-07} & 4.236\cdot 10^{+04} & 1.17\cdot 10^{-12} & 6.39\cdot 10^{-13} & 5.16\cdot 10^{-14} \\
& 100 & 248 & 1.141\cdot 10^{-03} & 1.017\cdot 10^{+08} & 3.59\cdot 10^{-18} & 1.48\cdot 10^{-18} & 3.74\cdot 10^{-20} \\
& 100 & 250 & 8.000\cdot 10^{-01} & 7.130\cdot 10^{+10} & 3.28\cdot 10^{-21} & 1.16\cdot 10^{-21} & 1.54\cdot 10^{-23} \\
& 100 & 252 & 1.250\cdot 10^{+02} & 1.114\cdot 10^{+13} & 1.84\cdot 10^{-22} & 6.12\cdot 10^{-23} & 6.26\cdot 10^{-25} \\
& 100 & 254 & 6.242\cdot 10^{-01} & 5.563\cdot 10^{+10} & 9.36\cdot 10^{-17} & 4.15\cdot 10^{-17} & 1.41\cdot 10^{-18} \\
& 100 & 256 & 3.308\cdot 10^{-04} & 2.948\cdot 10^{+07} & 8.08\cdot 10^{-11} & 4.84\cdot 10^{-11} & 5.77\cdot 10^{-12} \\
& 100 & 257 & 1.310\cdot 10^{+02} & 3.126\cdot 10^{+10} & 2.75\cdot 10^{-14} & 1.38\cdot 10^{-14} & 7.93\cdot 10^{-16} \\
& 100 & 258 & 1.172\cdot 10^{-11} & 1.045\cdot 10^{+00} & 9.47\cdot 10^{-01} & 9.46\cdot 10^{-01} & 9.41\cdot 10^{-01} \\
& 100 & 259 & 4.753\cdot 10^{-08} & 1.167\cdot 10^{+01} & 5.44\cdot 10^{-02} & 5.10\cdot 10^{-02} & 3.91\cdot 10^{-02} \\
& 100 & 260 & 1.268\cdot 10^{-10} & 1.130\cdot 10^{+01} & 6.33\cdot 10^{-02} & 5.95\cdot 10^{-02} & 4.62\cdot 10^{-02} \\ \hline
Md & 101 & 245 & 2.852\cdot 10^{-11} & 7.001\cdot 10^{-03} & 8.66\cdot 10^{+11} & 1.58\cdot 10^{+12} & 1.96\cdot 10^{+13} \\
& 101 & 247 & 6.338\cdot 10^{-09} & 1.556\cdot 10^{+00} & 1.73\cdot 10^{-01} & 1.67\cdot 10^{-01} & 1.42\cdot 10^{-01} \\
& 101 & 255 & 3.422\cdot 10^{-02} & 8.401\cdot 10^{+06} & 1.20\cdot 10^{-13} & 6.27\cdot 10^{-14} & 4.12\cdot 10^{-15} \\
& 101 & 257 & 6.338\cdot 10^{-09} & 1.546\cdot 10^{+07} & 1.37\cdot 10^{-12} & 7.52\cdot 10^{-13} & 6.16\cdot 10^{-14} \\
& 101 & 259 & 3.422\cdot 10^{-02} & 4.536\cdot 10^{+04} & 1.28\cdot 10^{-07} & 9.04\cdot 10^{-08} & 2.11\cdot 10^{-08} \\ \hline
\end{array}
$
\newpage
\textbf{Table 2a} (continued)

$
\begin{array}{|l|l|l|l|l|l|l|l|} \hline
& Z & A & T_{SF}(yr) & T_{SF}/T_{SF}^{C} & \Gamma _{\pi
^{-}}/\Gamma _{SF} & \Gamma _{\pi ^{0}}/\Gamma _{SF} & \Gamma _{\pi^{+}}/\Gamma _{SF} \\ \hline
No & 102 & 250 & 7.922\cdot 10^{-07} & 7.061\cdot 10^{-01} & 6.41\cdot 10^{+00} & 6.67\cdot 10^{+00} & 7.91\cdot 10^{+00}\\
& 102 & 251 & 3.169\cdot 10^{-07} & 7.779\cdot 10^{+01} & 3.33\cdot 10^{-09} & 2.17\cdot 10^{-09} &3.69\cdot 10^{-10} \\
& 102 & 252 & 2.852\cdot 10^{-07} & 2.542\cdot 10^{+04} & 9.39\cdot 10^{-18} & 3.96\cdot 10^{-18} & 1.05\cdot 10^{-19}\\
& 102 & 254 & 9.126\cdot 10^{-04} & 8.134\cdot 10^{+07} & 1.02\cdot 10^{-24} & 3.02\cdot 10^{-25} & 1.91\cdot 10^{-27}\\
& 102 & 256 & 1.711\cdot 10^{-05} & 1.525\cdot 10^{+06} & 3.58\cdot 10^{-16} & 1.64\cdot 10^{-16} & 6.30\cdot 10^{-18}\\
& 102 & 257 & 5.324\cdot 10^{-05} & 1.307\cdot 10^{+04} & 3.53\cdot 10^{-10} & 2.19\cdot 10^{-10} & 2.97\cdot 10^{-11}\\
& 102 & 258 & 3.803\cdot 10^{-11} & 3.389\cdot 10^{+00} & 7.47\cdot 10^{-02} & 7.06\cdot 10^{-02} & 5.58\cdot 10^{-02}\\
& 102 & 259 & 1.141\cdot 10^{-03} & 2.800\cdot 10^{+05} & 1.68\cdot 10^{-11} & 9.75\cdot 10^{-12} & 1.06\cdot 10^{-12}\\
& 102 & 260 & 3.359\cdot 10^{-09} & 2.994\cdot 10^{+02} & 2.62\cdot 10^{-05} & 2.08\cdot 10^{-05} & 7.96\cdot 10^{-06}\\
& 102 & 262 & 1.584\cdot 10^{-10} & 1.412\cdot 10^{+01} & 1.30\cdot 10^{-02} & 1.18\cdot 10^{-02} & 7.94\cdot 10^{-03}\\ \hline
Lr & 103 & 253 & 4.183\cdot 10^{-06} & 1.027\cdot 10^{+03} & 1.39\cdot 10^{-25} & 3.97\cdot 10^{-26} &2.14\cdot 10^{-28} \\
& 103 & 255 & 6.854\cdot 10^{-04} & 1.680\cdot 10^{+05} & 7.00\cdot 10^{-28} & 1.77\cdot 10^{-28} & 5.75\cdot 10^{-31} \\
& 103 & 257 & 6.274\cdot 10^{-05} & 1.540\cdot 10^{+04} & 1.25\cdot 10^{-16} & 5.59\cdot 10^{-17} & 1.96\cdot 10^{-18} \\
& 103 & 259 & 9.823\cdot 10^{-07} & 2.411\cdot 10^{+02} & 7.32\cdot 10^{-08} & 5.09\cdot 10^{-08} & 1.13\cdot 10^{-08} \\
& 103 & 261 & 7.415\cdot 10^{-05} & 1.820\cdot 10^{+04} & 2.85\cdot 10^{-11} & 1.67\cdot 10^{-11} & 1.84\cdot 10^{-12} \\ \hline
Rf & 104 & 253 & 1.521\cdot 10^{-12} & 3.734\cdot 10^{-04} & 1.16\cdot 10^{-64} & 4.57\cdot 10^{-66} &6.44\cdot 10^{-72} \\
& 104 & 254 & 7.288\cdot 10^{-13} & 6.496\cdot 10^{-02} & 8.87\cdot 10^{-68} & 2.96\cdot 10^{-69} & 2.11\cdot 10^{-75} \\
& 104 & 255 & 9.190\cdot 10^{-08} & 2.256\cdot 10^{+01} & 1.81\cdot 10^{-75} & 4.10\cdot 10^{-77} & 5.87\cdot 10^{-84} \\
& 104 & 256 & 1.965\cdot 10^{-10} & 1.751\cdot 10^{+01} & 7.98\cdot 10^{-24} & 2.47\cdot 10^{-24} & 1.91\cdot 10^{-26} \\
& 104 & 257 & 1.065\cdot 10^{-05} & 2.614\cdot 10^{+03} & 6.30\cdot 10^{-39} & 9.09\cdot 10^{-40} & 2.81\cdot 10^{-43} \\
& 104 & 258 & 4.436\cdot 10^{-10} & 3.954\cdot 10^{+01} & 1.67\cdot 10^{-13} & 8.74\cdot 10^{-14} & 5.93\cdot 10^{-15} \\
& 104 & 259 & 1.331\cdot 10^{-06} & 3.267\cdot 10^{+01} & 2.13\cdot 10^{-16} & 9.65\cdot 10^{-17} & 3.55\cdot 10^{-18} \\
& 104 & 260 & 6.338\cdot 10^{-10} & 5.648\cdot 10^{+01} & 1.12\cdot 10^{-09} & 7.16\cdot 10^{-10} & 1.06\cdot 10^{-10} \\
& 104 & 261 & 2.091\cdot 10^{-05} & 5.134\cdot 10^{+03} & 8.73\cdot 10^{-17} & 3.87\cdot 10^{-17} & 1.31\cdot 10^{-18} \\
& 104 & 262 & 6.654\cdot 10^{-08} & 5.931\cdot 10^{+03} & 6.72\cdot 10^{-15} & 3.27\cdot 10^{-15} & 1.62\cdot 10^{-16} \\ \hline
Db\pi F & 105 & 255 & 2.535\cdot 10^{-07} & 6.223\cdot 10^{+01} & 2.29\cdot 10^{+11} & 4.08\cdot 10^{+11}& 4.44\cdot 10^{+12}\\ \hline
Sg \pi F & 106 & 258 & 9.190\cdot 10^{-11} & 8.190\cdot 10^{+00} & 1.04\cdot 10^{+04} & 1.27\cdot 10^{+04} &2.96\cdot 10^{+04} \\
\pi F & 106 & 259 & 7.605\cdot 10^{-08} & 1.867\cdot 10^{+01} & 4.36\cdot 10^{+06} & 6.10\cdot 10^{+06} & 2.47\cdot 10^{+07}\\
\pi F & 106 & 260 & 2.218\cdot 10^{-10} & 1.977\cdot 10^{+01} & 2.12\cdot 10^{+08} & 3.24\cdot 10^{+08} & 1.86\cdot 10^{+09}\\
\pi F & 106 & 261 & 8.239\cdot 10^{-08} & 2.022\cdot 10^{+01} & 7.76\cdot 10^{+10} & 1.34\cdot 10^{+11} & 1.39\cdot 10^{+12}\\
\pi F & 106 & 263 & 8.556\cdot 10^{-08} & 2.100\cdot 10^{+01} & 1.31\cdot 10^{+27} & 5.17\cdot 10^{+27} & 1.54\cdot 10^{+30}\\
& 106 & 265 & 4.119\cdot 10^{-07} & 1.011\cdot 10^{+02} & 1.01\cdot 10^{-91} & 1.00\cdot 10^{-93} & 4.69\cdot 10^{-102} \\ \hline
Bh & 107 & 261 & 3.803\cdot 10^{-09} & 9.334\cdot 10^{-01} & 7.93\cdot 10^{-01} & 7.89\cdot 10^{-01} & 7.72\cdot 10^{-01} \\ \hline
Hs \pi F & 108 & 264 & 6.338\cdot 10^{-11} & 5.648\cdot 10^{+00} & 1.08\cdot 10^{+02} & 1.20\cdot 10^{+02} &1.84\cdot 10^{+02} \\
& 108 & 265 & 1.521\cdot 10^{-10} & 3.734\cdot 10^{-02} & 5.28\cdot 10^{-05} & 4.25\cdot 10^{-05} & 1.72\cdot 10^{-05} \\
& 108 & 267 & 3.169\cdot 10^{-09} & 7.779\cdot 10^{-01} & 3.84\cdot 10^{-01} & 3.76\cdot 10^{-01} & 3.44\cdot 10^{-01}\\ \hline
\end{array}
$
\newpage
\textbf{Table 2.b:} Theoretical predictions of the $\Gamma _{\pi ^{0,\pm }}/\Gamma
_{SF}$ yields for heavy parent nuclei with Z between Z=100 and Z=108,
obtained for $\alpha =0.75$ by using Eqs. (38)-(40) and $T_{SF}$ from Ref.[32].

\noindent $
\begin{array}{|l|l|l|l|l|l|l|l|} \hline
& Z & A & T_{SF} & \frac{T_{SF}}{T_{SF}^{C}} & \frac{\Gamma _{\pi ^{-}}%
}{\Gamma _{SF}} & \frac{\Gamma _{\pi ^{0}}}{\Gamma _{SF}} & \frac{\Gamma
_{\pi ^{+}}}{\Gamma _{SF}} \\ \hline
Cf&98&238&6.654\cdot 10^{-10}&5.931\cdot 10^{+01}&3.96\cdot 10^{-03}&4.47\cdot 10^{-03}&2.41\cdot 10^{-03}\\
&98&240&1.008\cdot 10^{-04}&8.981\cdot 10^{+06}&7.44\cdot 10^{-09}&1.11\cdot 10^{-08}&1.39\cdot 10^{-09}\\
&98&242&4.654\cdot 10^{-02}&4.148\cdot 10^{+09}&1.24\cdot 10^{-10}&2.03\cdot 10^{-10}&1.61\cdot 10^{-11}\\
&98&246&1.800\cdot 10^{+03}&1.604\cdot 10^{+14}&1.39\cdot 10^{-12}&2.50\cdot 10^{-12}&1.20\cdot 10^{-13}\\
&98&248&3.200\cdot 10^{+04}&2.852\cdot 10^{+15}&1.58\cdot 10^{-12}&2.83\cdot 10^{-12}&1.38\cdot 10^{-13}\\
&98&249&8.000\cdot 10^{+10}&1.964\cdot 10^{+19}&7.33\cdot 10^{-15}&1.47\cdot 10^{-14}&3.97\cdot 10^{-16}\\
&98&250&1.700\cdot 10^{+04}&1.515\cdot 10^{+15}&2.03\cdot 10^{-11}&3.46\cdot 10^{-11}&2.24\cdot 10^{-12}\\
&98&252&8.600\cdot 10^{+01}&7.665\cdot 10^{+12}&3.74\cdot 10^{-09}&5.69\cdot 10^{-09}&6.58\cdot 10^{-10}\\
&98&254&1.667\cdot 10^{-01}&1.486\cdot 10^{+10}&6.11\cdot 10^{-07}&8.32\cdot 10^{-07}&1.69\cdot 10^{-07}\\
&98&256&2.282\cdot 10^{-05}&2.033\cdot 10^{+06}&2.41\cdot 10^{-04}&2.88\cdot 10^{-04}&1.14\cdot 10^{-04}\\
&99&253&6.300\cdot 10^{+05}&1.546\cdot 10^{+14}&1.87\cdot 10^{-11}&3.18\cdot 10^{-11}&2.04\cdot 10^{-12}\\
&99&255&2.600\cdot 10^{+03}&6.382\cdot 10^{+11}&5.73\cdot 10^{-09}&8.62\cdot 10^{-09}&1.04\cdot 10^{-09}\\ \hline
Fm&100&242&2.535\cdot 10^{-11}&2.259\cdot 10^{+00}&1.39\cdot 10^{-01}&1.45\cdot 10^{-01}&1.17\cdot 10^{-01}\\
&100&244&1.046\cdot 10^{-10}&9.320\cdot 10^{+00}&1.53\cdot 10^{-02}&1.67\cdot 10^{-02}&1.05\cdot 10^{-02}\\
&100&245&1.255\cdot 10^{-04}&3.080\cdot 10^{+04}&2.74\cdot 10^{-08}&3.99\cdot 10^{-08}&5.76\cdot 10^{-09}\\
&100&246&4.753\cdot 10^{-07}&4.236\cdot 10^{+04}&8.12\cdot 10^{-08}&1.15\cdot 10^{-07}&1.88\cdot 10^{-08}\\
&100&248&1.141\cdot 10^{-03}&1.017\cdot 10^{+08}&3.91\cdot 10^{-11}&6.56\cdot 10^{-11}&4.57\cdot 10^{-12}\\
&100&250&8.000\cdot 10^{-01}&7.130\cdot 10^{+10}&5.41\cdot 10^{-13}&9.94\cdot 10^{-13}&4.31\cdot 10^{-14}\\
&100&252&1.250\cdot 10^{+02}&1.114\cdot 10^{+13}&8.65\cdot 10^{-14}&1.65\cdot 10^{-13}&5.84\cdot 10^{-15}\\
&100&254&6.242\cdot 10^{-01}&5.563\cdot 10^{+10}&2.14\cdot 10^{-10}&3.47\cdot 10^{-10}&2.92\cdot 10^{-11}\\
&100&256&3.308\cdot 10^{-04}&2.948\cdot 10^{+07}&7.79\cdot 10^{-07}&1.05\cdot 10^{-06}&2.21\cdot 10^{-07}\\
&100&257&1.310\cdot 10^{+02}&3.216\cdot 10^{+10}&6.01\cdot 10^{-09}&9.03\cdot 10^{-09}&1.10\cdot 10^{-09}\\
&100&258&1.172\cdot 10^{-11}&1.045\cdot 10^{+00}&9.67\cdot 10^{-01}&9.68\cdot 10^{-01}&9.64\cdot 10^{-01}\\
&100&259&4.753\cdot 10^{-08}&1.167\cdot 10^{+01}&1.70\cdot 10^{-01}&1.76\cdot 10^{-01}&1.45\cdot 10^{-01}\\
&100&260&1.268\cdot 10^{-10}&1.130\cdot 10^{+01}&1.86\cdot 10^{-01}&1.93\cdot 10^{-01}&1.60\cdot 10^{-01}\\ \hline
Md&101&245&2.852\cdot 10^{-11}&7.001\cdot 10^{-03}&1.02\cdot 10^{+07}&7.21\cdot 10^{+06}&4.33\cdot 10^{+07}\\
&101&247&6.338\cdot 10^{-09}&1.556\cdot 10^{+00}&3.56\cdot 10^{-01}&3.64\cdot 10^{-01}&3.24\cdot 10^{-01}\\
&101&255&3.422\cdot 10^{-02}&8.401\cdot 10^{+06}&1.84\cdot 10^{-08}&2.70\cdot 10^{-08}&3.73\cdot 10^{-09}\\
&101&257&6.297\cdot 10^{-02}&1.546\cdot 10^{+07}&7.43\cdot 10^{-08}&1.05\cdot 10^{-07}&1.70\cdot 10^{-08}\\
&101&259&1.848\cdot 10^{-04}&4.536\cdot 10^{+04}&6.98\cdot 10^{-05}&8.57\cdot 10^{-05}&2.96\cdot 10^{-05}\\ \hline
No&102&250&7.922\cdot 10^{-12}&7.061\cdot 10^{-01}&2.97\cdot 10^{+00}&2.90\cdot 10^{+00}&3.28\cdot 10^{+00}\\
&102&251&3.169\cdot 10^{-07}&7.779\cdot 10^{+01}&1.02\cdot 10^{-05}&1.31\cdot 10^{-05}&3.67\cdot 10^{-06}\\
&102&252&2.852\cdot 10^{-07}&2.542\cdot 10^{+04}&9.10\cdot 10^{-11}&1.49\cdot 10^{-10}&1.14\cdot 10^{-11}\\
&102&254&9.126\cdot 10^{-04}&8.134\cdot 10^{+07}&6.25\cdot 10^{-15}&1.26\cdot 10^{-14}&3.33\cdot 10^{-16}\\
&102&256&1.711\cdot 10^{-05}&1.525\cdot 10^{+06}&6.61\cdot 10^{-10}&1.04\cdot 10^{-09}&9.95\cdot 10^{-11}\\
&102&257&5.324\cdot 10^{-05}&1.307\cdot 10^{+04}&2.35\cdot 10^{-06}&3.11\cdot 10^{-06}&7.37\cdot 10^{-07}\\
&102&258&3.803\cdot 10^{-11}&3.389\cdot 10^{+00}&2.12\cdot 10^{-01}&2.20\cdot 10^{-01}&1.85\cdot 10^{-01}\\
&102&259&1.141\cdot 10^{-03}&2.800\cdot 10^{+05}&3.64\cdot 10^{-07}&5.01\cdot 10^{-07}&9.64\cdot 10^{-08}\\
&102&260&3.359\cdot 10^{-09}&2.994\cdot 10^{+02}&1.80\cdot 10^{-03}&2.07\cdot 10^{-03}&1.02\cdot 10^{-03}\\
&102&262&1.584\cdot 10^{-10}&1.412\cdot 10^{+01}&7.37\cdot 10^{-02}&7.79\cdot 10^{-02}&5.83\cdot 10^{-02}\\ \hline
\end{array}
$
\newpage
\textbf{Table 2.b} (continued)

\noindent $
\begin{array}{|l|l|l|l|l|l|l|l|} \hline
& Z & A & T_{SF} & \frac{T_{SF}}{T_{SF}^{C}} & \frac{\Gamma _{\pi ^{-}}%
}{\Gamma _{SF}} & \frac{\Gamma _{\pi ^{0}}}{\Gamma _{SF}} & \frac{\Gamma
_{\pi ^{+}}}{\Gamma _{SF}} \\ \hline
Lr&103&253&4.183\cdot 10^{-06}&1.027\cdot 10^{+03}&2.89\cdot 10^{-15}&5.94\cdot 10^{-15}&1.44\cdot 10^{-16}\\
&103&255&6.845\cdot 10^{-04}&1.680\cdot 10^{+05}&1.12\cdot 10^{-16}&2.47\cdot 10^{-16}&4.17\cdot 10^{-18}\\
&103&257&6.274\cdot 10^{-05}&1.540\cdot 10^{+04}&4.18\cdot 10^{-10}&6.66\cdot 10^{-10}&6.04\cdot 10^{-11}\\
&103&259&9.823\cdot 10^{-07}&2.411\cdot 10^{+02}&5.96\cdot 10^{-05}&7.35\cdot 10^{-05}&2.49\cdot 10^{-05}\\
&103&261&7.415\cdot 10^{-05}&1.820\cdot 10^{+04}&5.41\cdot 10^{-07}&7.39\cdot 10^{-07}&1.48\cdot 10^{-07}\\ \hline
Rf&104&253&1.521\cdot 10^{-12}&3.734\cdot 10^{-04}&9.87\cdot 10^{-38}&6.18\cdot 10^{-37}&4.77\cdot 10^{-41}\\
&104&254&7.288\cdot 10^{-13}&6.496\cdot 10^{-02}&1.26\cdot 10^{-39}&8.72\cdot 10^{-39}&4.15\cdot 10^{-43}\\
&104&255&9.190\cdot 10^{-08}&2.256\cdot 10^{+01}&3.53\cdot 10^{-44}&3.04\cdot 10^{-43}&4.51\cdot 10^{-48}\\
&104&256&1.965\cdot 10^{-10}&1.751\cdot 10^{+01}&3.49\cdot 10^{-14}&6.80\cdot 10^{-14}&2.17\cdot 10^{-15}\\
&104&257&1.065\cdot 10^{-05}&2.614\cdot 10^{+03}&4.98\cdot 10^{-23}&1.50\cdot 10^{-22}&5.00\cdot 10^{-25}\\
&104&258&4.436\cdot 10^{-10}&3.954\cdot 10^{+01}&3.34\cdot 10^{-08}&4.85\cdot 10^{-08}&7.16\cdot 10^{-09}\\
&104&259&1.331\cdot 10^{-06}&3.267\cdot 10^{+02}&6.50\cdot 10^{-10}&1.02\cdot 10^{-09}&9.77\cdot 10^{-11}\\
&104&260&6.338\cdot 10^{-10}&5.648\cdot 10^{+01}&5.54\cdot 10^{-06}&7.19\cdot 10^{-06}&1.87\cdot 10^{-06}\\
&104&261&2.091\cdot 10^{-05}&5.134\cdot 10^{+03}&3.53\cdot 10^{-10}&5.64\cdot 10^{-10}&5.02\cdot 10^{-11}\\
&104&262&6.654\cdot 10^{-08}&5.931\cdot 10^{+03}&4.38\cdot 10^{-09}&6.63\cdot 10^{-09}&7.82\cdot 10^{-10}\\ \hline
Db&105&255&2.535\cdot 10^{-07}&6.223\cdot 10^{+01}&3.41\cdot 10^{+06}&2.47\cdot 10^{+06}&1.31\cdot 10^{+07}\\
&105&257&2.535\cdot 10^{-07}&6.223\cdot 10^{+01}&1.09\cdot 10^{+12}&6.00\cdot 10^{+11}&1.30\cdot 10^{+13}\\
&105&261&3.169\cdot 10^{-07}&7.779\cdot 10^{+01}&2.93\cdot 10^{-20}&7.72\cdot 10^{-20}&5.21\cdot 10^{-22}\\
&105&263&1.521\cdot 10^{-06}&3.734\cdot 10^{+02}&1.44\cdot 10^{-12}&2.60\cdot 10^{-12}&1.26\cdot 10^{-13}\\ \hline
Sg&106&258&9.190\cdot 10^{-11}&8.190\cdot 10^{+00}&1.99\cdot 10^{+02}&1.78\cdot 10^{+02}&3.21\cdot 10^{+02}\\
&106&259&7.605\cdot 10^{-08}&1.867\cdot 10^{+01}&6.47\cdot 10^{+03}&5.35\cdot 10^{+03}&1.42\cdot 10^{+04}\\
&106&260&2.218\cdot 10^{-10}&1.977\cdot 10^{+01}&6.17\cdot 10^{+04}&4.86\cdot 10^{+04}&1.65\cdot 10^{+05}\\
&106&261&8.239\cdot 10^{-08}&2.022\cdot 10^{+01}&1.89\cdot 10^{+06}&1.39\cdot 10^{+06}&6.93\cdot 10^{+06}\\
&106&263&8.556\cdot 10^{-08}&2.100\cdot 10^{+01}&5.03\cdot 10^{+15}&2.31\cdot 10^{+15}&1.28\cdot 10^{+17}\\
&106&265&4.119\cdot 10^{-07}&1.011\cdot 10^{+02}&1.22\cdot 10^{-53}&1.69\cdot 10^{-52}&2.22\cdot 10^{-58}\\ \hline
Bh&107&261&3.803\cdot 10^{-09}&9.334\cdot 10^{-01}&8.76\cdot 10^{-01}&8.78\cdot 10^{-01}&8.66\cdot 10^{-01}\\ \hline
Hs&108&264&6.338\cdot 10^{-11}&5.648\cdot 10^{+00}&1.42\cdot 10^{+01}&1.34\cdot 10^{+01}&1.81\cdot 10^{+01}\\
&108&265&1.521\cdot 10^{-10}&3.734\cdot 10^{-02}&3.68\cdot 10^{-03}&4.15\cdot 10^{-03}&2.22\cdot 10^{-03}\\
&108&267&3.169\cdot 10^{-09}&7.779\cdot 10^{-01}&5.78\cdot 10^{-01}&5.85\cdot 10^{-01}&5.51\cdot 10^{-01}\\ \hline
\end{array}
$

\newpage
\textbf{Table 3}: Experimental results on spontaneous NPIR-yields

\noindent
\hspace{-1.5cm} \begin{tabular}{|l|l|l|l|l|l|} \hline
Year &  \quad \quad \quad Authors & 
\quad \quad Laboratory & Parent  & $\Gamma _{\pi
^{0}}/\Gamma _{SF}$ & $\Gamma _{\pi ^{\pm }}/\Gamma _{SF}$ \\
&&&nuclei&&\\ \hline
1986 & D.B.Ion et al.  & IFIN-Bucharest 
& $^{259}$Md &  & $<10^{-5}$ \\
&Rev.Roum.Phys.\textbf{31},551&Romania&&&\\ \hline
1987 & D.Bucurescu et al. & IFIN-
Romania & $^{252}$Cf &  & $<10^{-8}$ \\
&Rev.Roum.Phys.\textbf{32}, 849 &Bucharest&&&\\ \hline
1988 & M.Ivascu et al.& IFIN-Bucharest 
& $^{252}$Cf &  & $<10^{-8}$ \\
&Rev.Roum.Phys.\textbf{32}, 937 &Romania&&&\\ \hline
1988 & C.Cerruti et al. & CEN-Saclay  & $%
^{252}$Cf & $<10^{-8}$ &  \\
&Z.Phys. A$\mathbf{329}$, 383 &France&&&\\ \hline
1988 & J.R.Beene et al. & ORNL USA & $^{252}$Cf
& $<1.5\cdot 10^{-9}$ &  \\
&Phys.Rev. \textbf{C38}, 569 &&&&\\ \hline
1989 & J. Julien et al. & CEN-Saclay France & $%
^{252}$Cf & $<10^{-12}$ &  \\
&Z.Phys, \textbf{A332}, 473 &&&&\\ \hline
1989 & Yu.Adamchuk et al. & I.V.Kurhatov
Russia & $^{252}$Cf &  & $<5\cdot 10^{-8}$ \\
&Sov. J. Nucl. Phys. \textbf{49},932 &&&&\\ \hline
1989 & D.B.Ion et al. & IFIN-Bucharest
 & $^{257}$Fm &  & $(1.2\pm 0.2)$ \\
&Rev.Roum.Phys. \textbf{34}, 261 &Romania&&&$\cdot 10^{-3}(\pi ^{-})$\\ \hline
1989 & S.Stanislaus et al. & University British
& $^{252}$Cf & $<3.3\cdot 10^{-10}$ &  \\
&Phys.Rev.\textbf{\ C39}, 295 &Columbia Canada &&&\\ \hline
1989 & S.Stanislaus et al. & University British
& $^{238}$U & $<3.1\cdot 10^{-4}$ &  \\
&Phys.Rev. \textbf{C39}, 295 &Columbia Canada &&&\\ \hline
1991 & J.N.Knudson et al. & LANL USA & $^{252}$%
Cf & $<1.37\cdot 10^{-11}$ &  \\
&Phys.Rev. \textbf{C44}, 2869 &&&&\\ \hline
1991 & V.Bellini et al. [19] Proc. & CEN Saclay France
& $^{252}$Cf & $<3 \cdot10^{-13}$ &  \\
&Bratislava Conference &&&&\\ \hline
1991 & K.Janko et al. [22] Proc. & Comenius University
 & $^{257}$Fm &  & $<7\cdot 10^{-5}$ \\
&Bratislava Conference &Czechoslovakia&&&$(\pi ^{+})$\\ \hline
1992 & H.Otsu et al.  & Tokio University  & $%
^{252}$Cf &  & $<1.3\cdot 10^{-8}$ \\
&Z.Phys. \textbf{A 342}, 483&Japan&&&$(\pi ^{-})$\\ \hline
2002 & Khryachkov et al. & LIPPE Obninsk  & $^{252}$Cf &  & Not \\
&Instr. Exp. Tech. \textbf{45}, 615 &Russia&&&estimated\\ \hline
\end{tabular}

\begin{figure}
 \begin{center}
  \hspace{-2cm} \includegraphics[width=14cm]{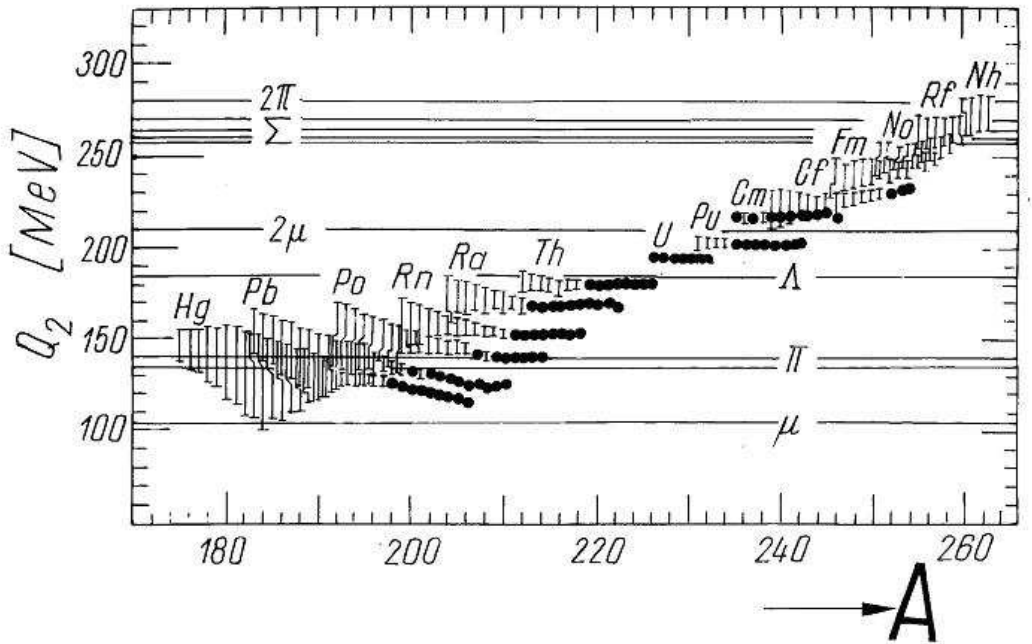}
 \end{center}
 \caption{The energies $Q_{\pi }=Q_{\pi F}$ calculated by using Eq.
(3) and Tables of Wapstra and Audi.}
 \label{fg1}
\end{figure}

\begin{figure}
\hspace{-5mm}
\includegraphics[width=6.5cm]{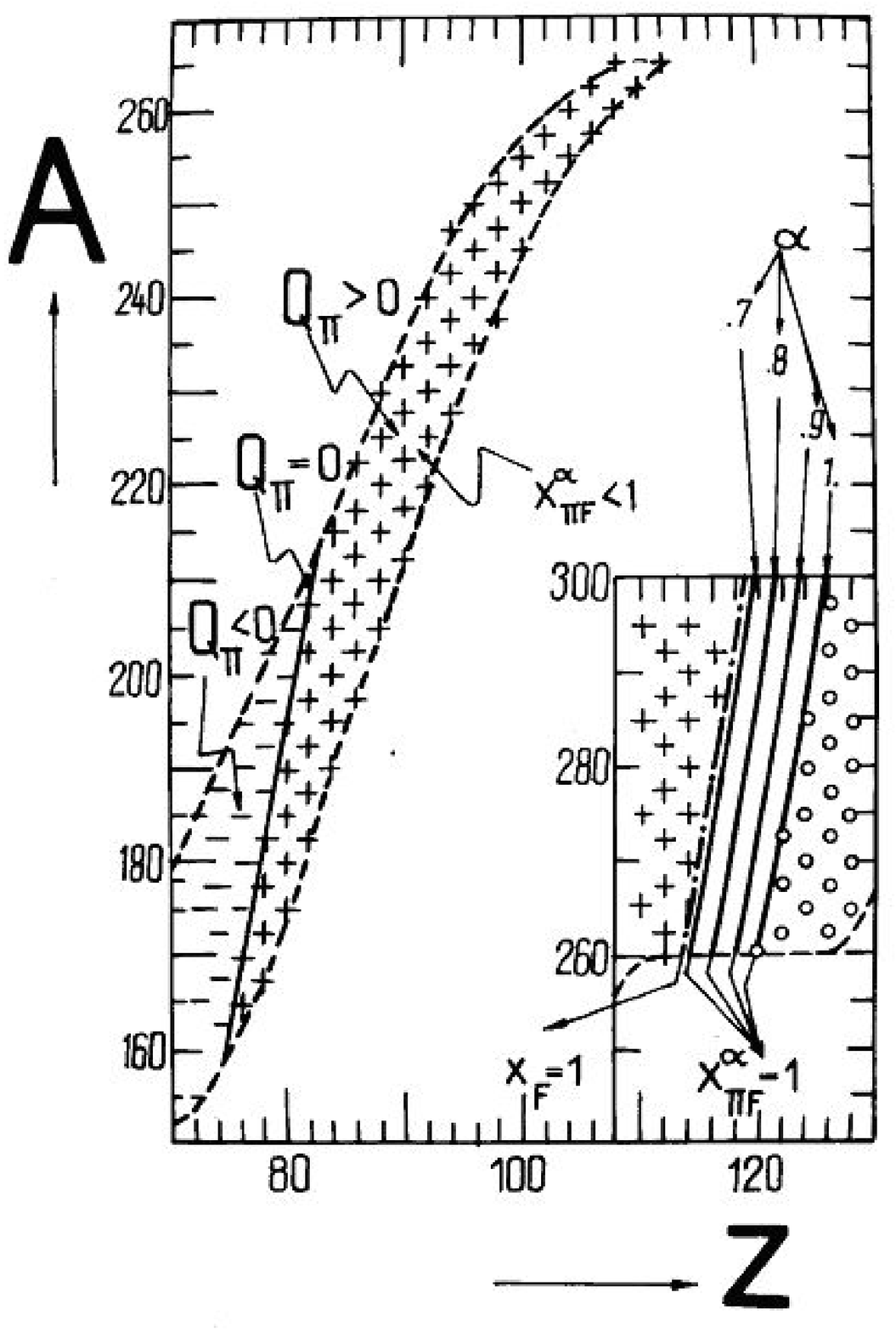}
\hspace{5mm}
 \includegraphics[width=6cm]{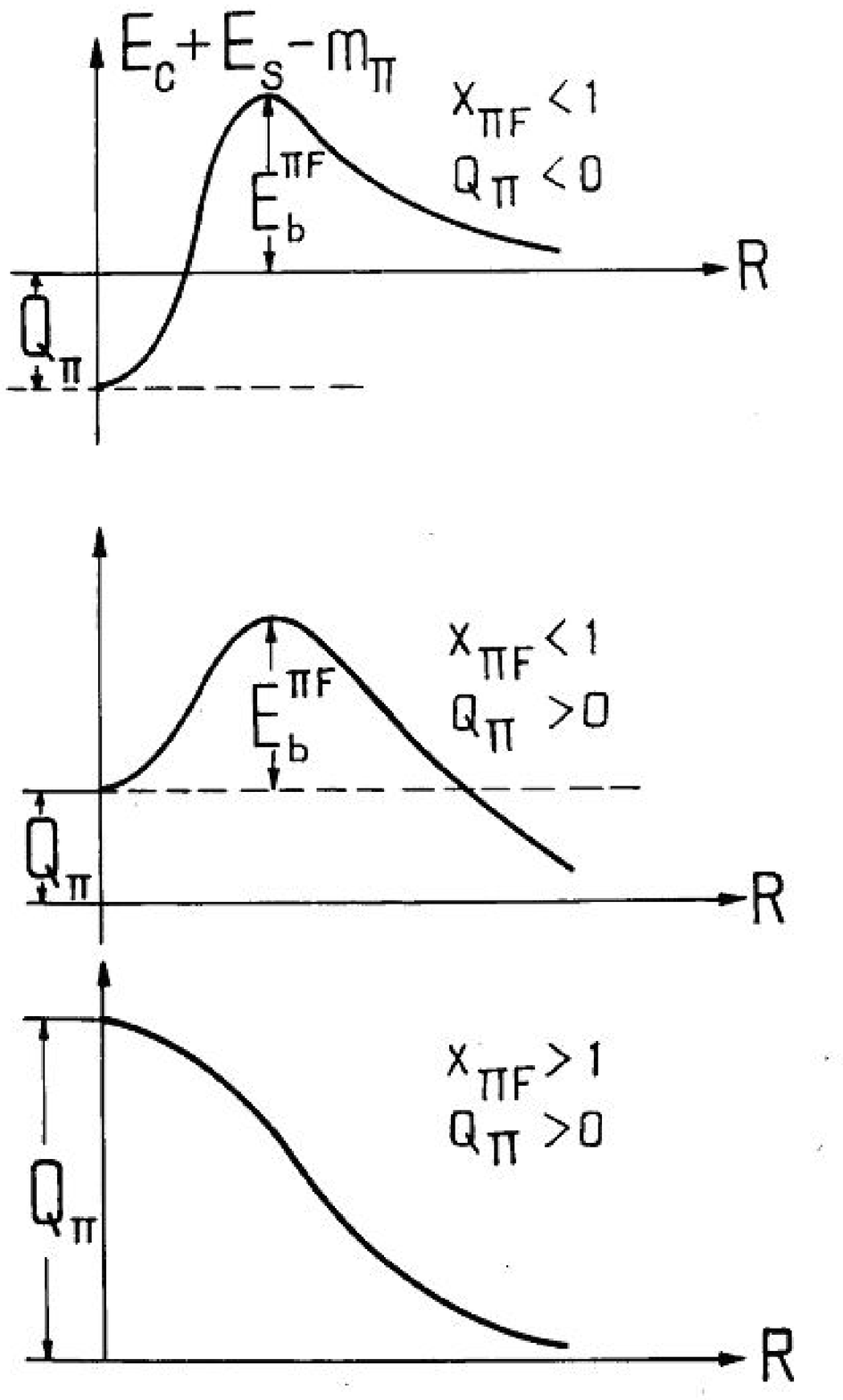}
\parbox{60mm}{
\caption{The physical regions from the plane (A,Z) in which the
parent nuclei are able to emit pions by tunnelling pi-fission barr\^{i}er
(++++) and without tunnelling $\pi F-$barrier (oooo) (see the text).}}
\hspace{5mm} \parbox[top]{60mm}{
\caption{Definition of $\pi F-$barrier as a function of distance
between the fragments during the pionic fission.}}
\label{fg2}
\end{figure}

\begin{figure}
 \begin{center}
  \hspace{-2cm} \includegraphics[width=10cm]{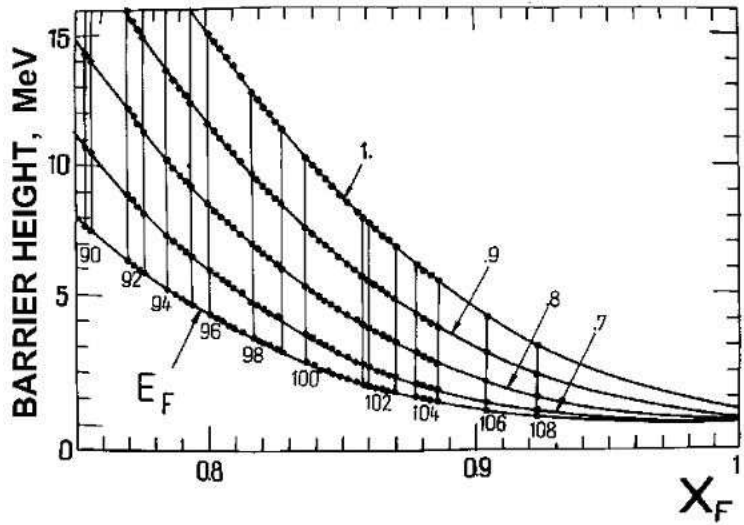}
 \end{center}
 \caption{The values of the barrier height E$^{\pi F}\left( LD\right) $
for the neutral pion emission during spontaneous fission fission.}
 \label{fg4}
\end{figure}

\begin{figure}
 \begin{center}
  \hspace{-2cm} \includegraphics[width=7cm]{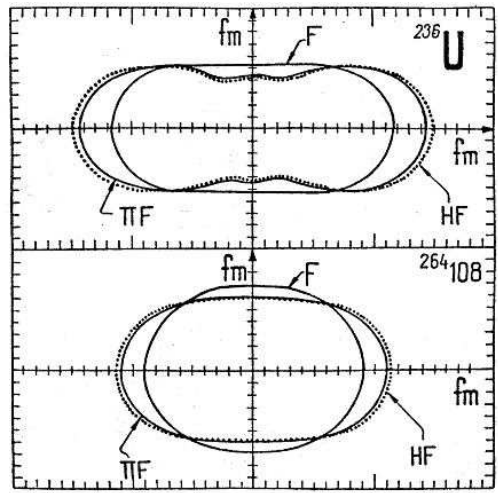}
 \end{center}
 \caption{The nuclear configuration at the saddle point for spontaneous
fission (F), spontaneous hyperfission (HF) and spontaneous pionic fission ($%
\pi F$).}
 \label{fg5}
\end{figure}

\begin{figure}
 \begin{center}
  \hspace{-2cm} \includegraphics[width=10cm]{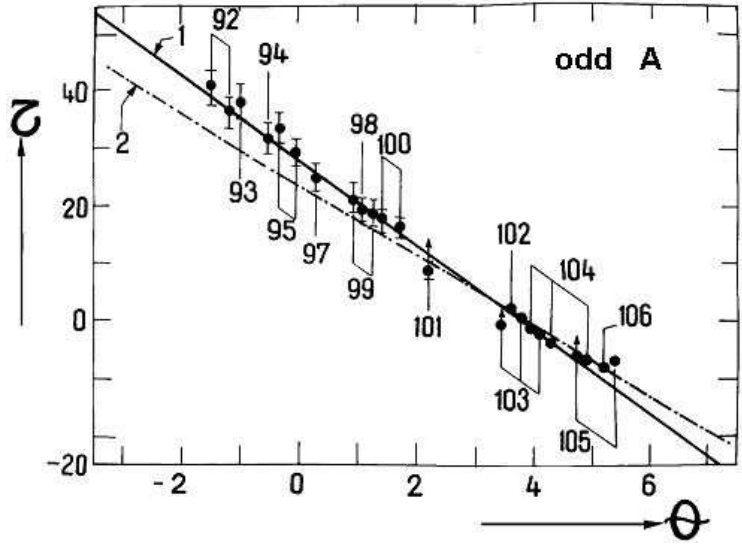}
 \end{center}
 \caption{The experimental values of $\tau (\theta )$are compared wit
the linear fit (35).}
 \label{fg6}
\end{figure}

\begin{figure}
 \begin{center}
  \hspace{-2cm} \includegraphics[width=9cm]{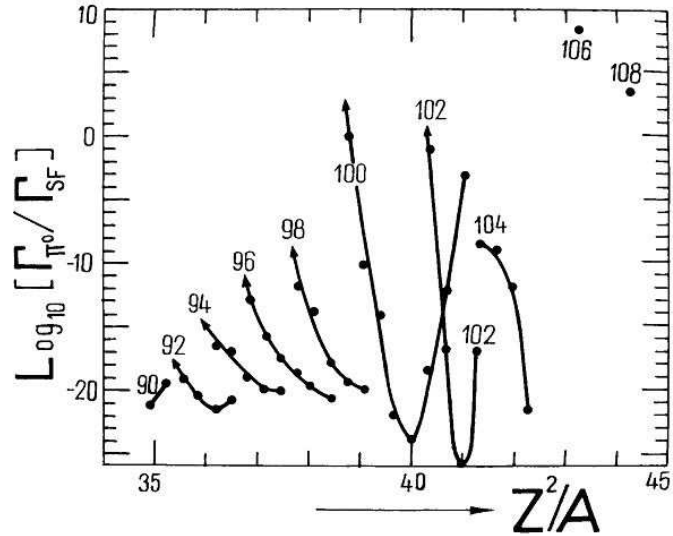}
 \end{center}
 \caption{The predicted values of \ $\frac{\Gamma _{\pi 0}}{\Gamma _{SF}%
}$\ calculated with (36) for $\alpha =1$ , for eve-even parent nuclei.}
 \label{fg7}
\end{figure}

\begin{figure}
 \begin{center}
  \hspace{-2cm} \includegraphics[width=12cm]{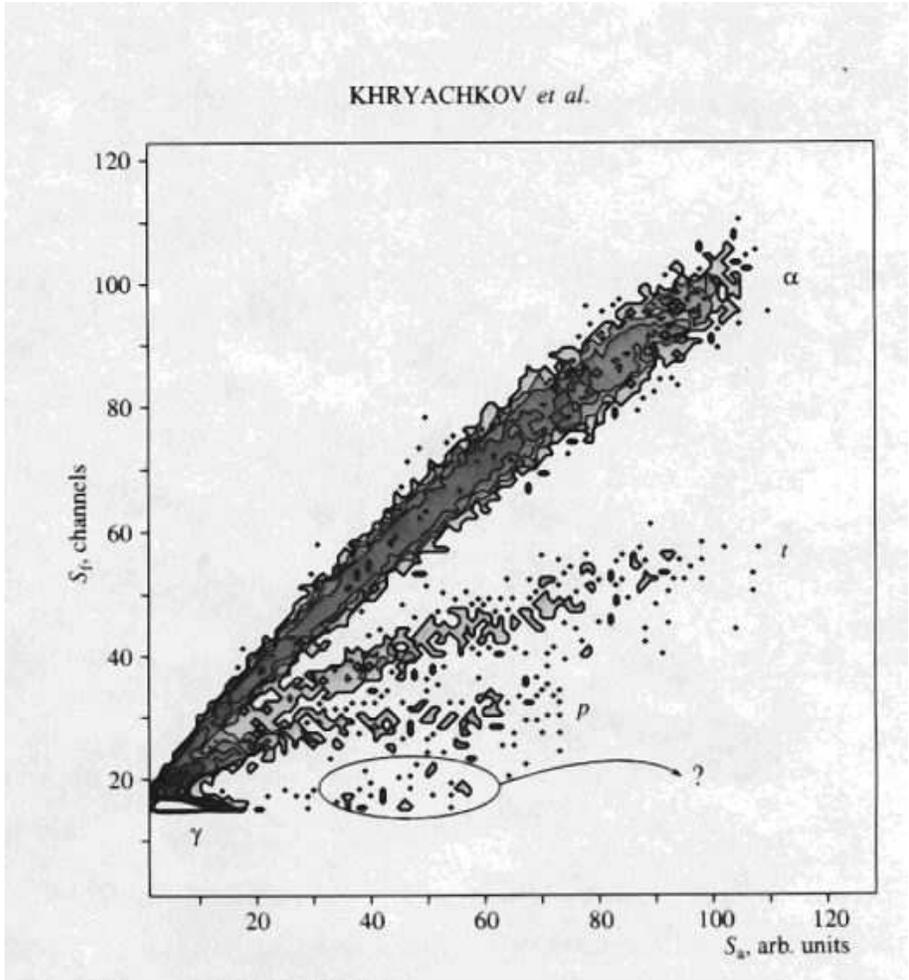}
 \end{center}
 \caption{The experimental results (see Fig. 9 in Ref. [27]) on the
light charged particle emission during the spontaneous fission of $^{252}Cf$.}
 \label{fg8}
\end{figure}

\end{document}